\documentclass[12pt]{article}
\usepackage{epsfig, amsmath, amssymb}
\usepackage{graphicx,psfrag} 
\newcommand{\nn}{\nonumber}
\newcommand{\be}{\begin{equation}}
\newcommand{\ee}{\end{equation}}
\newcommand{\ben}{\begin{equation}}
\newcommand{\een}{\end{equation}}
\newcommand{\bea}{\begin{eqnarray}}
\newcommand{\eea}{\end{eqnarray}}
\newcommand{\bA}{\begin{array}}
\newcommand{\eA}{\end{array}}
\newcommand{\bc}{\begin{center}}
\newcommand{\ec}{\end{center}}
\newcommand{\al}{\alpha}

\newcommand{\ra}{\rightarrow}
\newcommand{\del}{\partial}

\newcommand{\ie}{{\it i.e.}}
\newcommand{\eg}{{\it e.g.}}

\begin{document}


\begin{titlepage}

\bc

\hfill 
\\         [25mm]

\vspace{2mm}

{\Huge Cosmological singularities and
 \\ [2mm] 2-dimensional dilaton gravity} 
\vspace{16mm}

{\large Ritabrata Bhattacharya,\ K.~Narayan,\ Partha Paul} \\
\vspace{3mm}
{\small \it Chennai Mathematical Institute, \\}
{\small \it SIPCOT IT Park, Siruseri 603103, India.\\}

\ec
\vspace{35mm}

\begin{abstract}
  We study Big-Bang or -Crunch cosmological singularities in
  2-dimensional dilaton-gravity-scalar theories, in general obtained
  by dimensional reduction of higher dimensional theories. The dilaton
  potential encodes information about the asymptotic data defining the
  theories, and encompasses various families such as flat space,
  $AdS$, conformally $AdS$ as arising from nonconformal branes, and
  more general nonrelativistic theories. We find a kind of universal
  near singularity behaviour independent of the dilaton potential,
  giving universal interrelations between the exponents defining the
  time behaviour near the cosmological singularity. More detailed
  analysis using a scaling ansatz enables finding various classes of
  cosmological backgrounds, recovering known examples such as the
  $AdS$ Kasner singularity as well finding as new ones. We give some
  comments on the dual field theory from this point of view.
\end{abstract}

\end{titlepage}

{\tiny 
\begin{tableofcontents}
\end{tableofcontents}
}


\section{Introduction}

It is a fond hope that string theory sheds light on Big Bang or Crunch
singularities and the early universe: there is a rich history of
previous investigations,
\eg\ \cite{Horowitz:1989bv}-\cite{Choudhury:2020yaa} (see also
\eg\ \cite{Craps:2006yb,Burgess:2011fa} for reviews).  Some of these
involve the worldsheet theory of strings propagating in the background
of cosmological singularities, the intuition being that the extended
nature of the string (its oscillatory modes and interactions thereof)
smoothes out the near singularity region. Others involve holography
\cite{Maldacena:1997re}-\cite{Aharony:1999ti}, the hope being that the
dual field theory is well-defined even if the bulk gravitational
theory breaks down in the vicinity of the singularity. For instance,
time-dependent deformations of $AdS/CFT$ were studied in
\cite{Das:2006dz}-\cite{Awad:2008jf}: the bulk develops a Big-Crunch
singularity and breaks down while the dual field theory subjected to
severe time-dependence (in some cases due to a severe time-varying
gauge coupling) may be hoped to march on: we will discuss these more
later. In many cases however, the analysis, while interesting and
instructive, suggests a singular response in the field theory as well.
Further insights on some of these were obtained in
\cite{Engelhardt:2014mea}-\cite{Engelhardt:2016kqb} in part via
different holographic screens. It would be desirable to have more
conclusive insights into possible resolutions of cosmological
singularities in string theory.

In this paper, we study time-dependent backgrounds in 2-dimensional
dilaton-gravity-matter theories, many of which are obtained by
dimensional reduction from higher dimensional theories. The 2-dim
theories here contain the 2-dim dilaton coupling to gravity as well as
an extra scalar and a dilaton potential. Some of the cosmological
singularities known in flat space and in the $AdS/CFT$ deformations
mentioned above are encompassed in these 2-dim theories: although the
reduction ansatz precludes very anisotropic backgrounds, it still
allows interesting structure. This recasting into 2-dimensions helps
organize the equations of motion governing such Big-Bang or -Crunch
singularities: we will often find it convenient to think in terms of a
Crunch. In particular we find a kind of ``universal near singularity''
behaviour near the (spacelike) Big-Crunch, defined by a region where
the fields are rapidly varying in time leading to large time
derivatives that eventually diverge: in this region the precise form
of the dilaton potential turns out to essentially disappear, thus
becoming irrelevant. This suggests a scaling ansatz for the pertinent
2-dimensional fields, \ie\ the metric function $e^f$, dilaton $X$ and
scalar $e^\Phi$, which is of power-law form in both time $t$ and space
$r$ variables. The universal near singularity behaviour above implies
universal relations between the exponents governing the time behaviour
near the singularity: in particular these relations are satisfied by
the flat space Kasner singularity that is ``mostly isotropic''.

A more detailed analysis of the 2-dimensional equations using this
scaling ansatz confirms this ``cosmological attractor'' type behaviour
and also fixes the precise values of all the exponents. In particular
this recovers the familiar flat space Kasner and $AdS$ Kasner Big Bang
singularities: as well as various familiar time-independent
backgrounds such as $AdS$, conformally $AdS$ and nonrelativistic
theories. Performing this analysis for more general dilaton potentials
obtained by reductions of certain classes of nonrelativistic theories
in fact reveals new families of time-dependent backgrounds with
cosmological singularities and hyperscaling violating Lifshitz
asymptotics far from the singular region. In JT gravity, this reveals
a cosmological background exhibiting Crunch behaviour where the
dilaton vanishes.

In sec.2, we briefly review certain aspects of the investigations in
\cite{Das:2006dz,Das:2006pw,Awad:2007fj,Awad:2008jf} on $AdS/CFT$ and
Big-Bang singularities. Sec.3 describes the effective 2-dimensional
dilaton-gravity-matter theories that encode such time-dependent
backgrounds, in particular those arising as reductions from higher
dimensions. Sec.4 discusses more detailed analysis on solving the
2-dim theory for Big-Bang or -Crunch singularities, in various
theories. Sec.5 has some comments on the holographic duals in the
$AdS/CFT$ cosmological singularities and we conclude finally with a
Discussion in Sec.6.

\section{$AdS/CFT$ and Big-Bang/Crunch singularities}

Time-dependent non-normalizable deformations of $AdS/CFT$ were studied
in \cite{Das:2006dz,Das:2006pw,Awad:2007fj,Awad:2008jf} towards
gaining insights via gauge/gravity duality into cosmological (Big-Bang
or -Crunch) singularities: some of these were further investigated in
\cite{Engelhardt:2014mea,Engelhardt:2015gta,Engelhardt:2015gla,
  Engelhardt:2016kqb}. The bulk gravity theory exhibits a cosmological
Big-Crunch (or -Bang) singularity and breaks down while the
holographic dual field theory (in the $AdS_5$ case) subject to a
severe time-dependent gauge coupling $g_{YM}^2=e^\Phi$ may be hoped to
provide insight into the dual dynamics: in this case the scalar $\Phi$
controls the gauge/string coupling.  There is a large family of such
backgrounds exhibiting cosmological singularities (we will discuss
them further in sec.5).  For instance, we obtain $AdS$-Kasner theories
as
\be\label{adsKasner}
ds^2 = {R_{AdS}^2\over r^2} (-dt^2 + \sum_i t^{2p_i} dx_i^2 + dr^2) , \quad
e^\Phi=t^\al\ ;\qquad \sum_i p_i=1\ ,\quad \sum_i p_i^2 = 1 - {1\over 2}\al^2\ .
\ee
The Kasner exponents $p_i$ are constrained as above. We thus see that
for constant (undeformed) scalar $\Phi$ with $\al=0$, the Kasner space
is necessarily anisotropic: the $p_i$ cannot all be equal. In this case,
the gauge theory lives on a time-dependent space but the gauge coupling
is not time-dependent.  An isotropic $AdS$-Kasner singularity
\be\label{ads5bb}
ds^2 = {R_{AdS}^2\over r^2} (-dt^2 + t^{2/3} dx_i^2 + dr^2)\ , \qquad\ \
e^\Phi=t^{2/\sqrt{3}}\ ,
\ee
has all Kasner exponents equal: this requires a nontrivial time-dependent
dilaton source $\Phi$ as well.
More general backgrounds can also be found involving $AdS$-FRW and
$AdS$-BKL spacetimes \cite{Awad:2007fj,Awad:2008jf}: these all have
spacelike singularities. There are also backgrounds with null
singularities: in these cases, the dual appears to be weakly coupled
gauge theory \cite{Das:2006pw}, modulo some caveats (sec.5).\ 
Similar backgrounds arise for $AdS_4\times X^7$ deformations although
the details differ slightly: \eg\ the isotropic $AdS_4$-Kasner
background has $p_i={1\over 2}$ and ${\al^2\over 2}={1\over 2}$ giving
\be\label{ads4bb}
ds^2 = {R_{AdS}^2\over r^2} (-dt^2 + t\,dx_i^2 + dr^2)\ , \qquad\ \
e^\Phi=t\ .
\ee
These also contain a Big-Crunch singularity as for the $AdS_5$ case,
but with different exponents controlling the approach to the
singularity (and a nontrivial $\Phi$ necessarily). In this case the
scalar $\Phi$ arises from the 4-form flux after compactification on
some 7-manifold $X^7$: see \cite{Balasubramanian:2010uk} for more
details on the 11-dim supergravity constructions of these backgrounds.

The above backgrounds are solutions to Einstein-dilaton gravity with
negative cosmological constant $\Lambda<0$: these arise as consistent
truncations of IIB string theory or M-theory on $AdS\times S$ spaces,
the flux giving rise to the effective cosmological constant.

We will refer to the scalar $\Phi$ as the ``string dilaton'' here
(although it is only in the $AdS_5/CFT_4$ case that $\Phi$ controls
the string coupling). The word \emph{dilaton} will refer to the scalar
that plays the role of the 2-dim dilaton, arising from Kaluza-Klein
compactification from higher dimensions (this controls the transverse
area of the higher dimensional space).

\section{Redux to 2d dilaton gravity: Big-Crunch attractors}

In what follows, we would like to explore Kaluza-Klein reductions of
these models of Big-Bang/Crunch singularities down to two bulk
dimensions. See \eg\ \cite{Strominger:1994tn,Nojiri:2000ja,
  Grumiller:2001ea,Grumiller:2002nm}
for a review of 2-dim theories obtained from dimensional reduction
(and \cite{Mann:1992yq}-\cite{Nojiri:2020tph} for some previous work on
2d cosmologies in part motivated by the CGHS model of 2-dim black holes
\cite{Callan:1992rs}).
2-dim dilaton gravity has been revisited in \cite{Almheiri:2014cka} in
the context of $AdS_2$ holography.
We are considering higher dimensional theories of Einstein
gravity with a scalar $\Phi$ and a potential $V$, the action being of
the form
\be\label{actionDdim}
S = {1\over 16 \pi G_D} \int d^Dx \sqrt{-g^{(D)}}\,
\Big( {\cal R} - {1\over 2} (\del\Phi)^2 - V \Big)\ .
\ee
We allow the potential $V$ to also contain metric data, \ie\ it is a
function $V(g,\Phi)$.  These sorts of theories (which we will expand
on later) were considered in \cite{Kolekar:2018sba,Kolekar:2018chf}
whose conventions we are following.

Let us take the reduction ansatz on some compact space $M^{D-2}$ to be
\be\label{ads2Redux1}
ds^2 \equiv\ g^{(2)}_{\mu\nu} dx^\mu dx^\nu + X^{4\over D-2} d\sigma_{D-2}^2\ ,
\qquad\qquad D=d_i+2\ .
\ee
It is convenient to define $d_i=D-2$ as the ``transverse'' part of the
bulk spacetime, apart from the 2-dimensional piece (which will contain
time $t$): in theories with holographic duals, $d_i$ will become the
boundary spatial dimension.  With this form of the 2-dim dilaton $X$,
the 2-dim dilaton gravity action is of the form\ $\int (X^2 {\cal R}^{(2)}
+ \ldots)$:\ also the dilaton $X^2$ is essentially the transverse area
$g_{ii}^{D-2}=X^2$. This ansatz for $M^{D-2}$ being a torus $T^{D-2}$
implies translation/rotational invariance in the spatial directions in
the higher dimensional background (as is the case for various
relativistic backgrounds such as $AdS$ or conformally $AdS$, or
nonrelativistic ones such as hyperscaling violating Lifshitz).

In detail, the reduction ansatz (\ref{ads2Redux1}) gives
\be\label{actionKEdil}
S = {1\over 16\pi G_2} \int d^2x \sqrt{-g^{(2)}}
\left(X^2 {\cal R}^{(2)} + {D-3\over D-2}\, {(\del X^2)^2 \over X^2} 
- {1\over 2} X^2 (\del\Phi)^2 - V X^2 \right)\ .
\ee
A total derivative term that arises in this reduction cancels with
a corresponding term from reduction of the Gibbons Hawking boundary term.
In \cite{Kolekar:2018sba,Kolekar:2018chf}, as in \cite{Almheiri:2014cka},
an additional Weyl transformation was performed to absorb the dilaton
kinetic term in the 2-dim Ricci scalar: note that a Weyl
transform acts as
\be\label{weyl0}
\sqrt{-g^{(2)}} \big[\Psi^2 {\cal R}^{(2)} + \lambda (\del\Psi)^2\big]\ \ \
\xrightarrow{\ g_{ab}^{(2)} = \Psi^{-\al/2} g_{ab}\ }\ \ \
\sqrt{-g} \big[\Psi^2 {\cal R} + (\lambda-\al) (\del\Psi)^2\big]\ 
\ee
Thus defining
\be\label{weyl}
g_{\mu\nu}=X^{{2(D-3)\over D-2}} g^{(2)}_{\mu\nu}\ ,
\ee
removes the kinetic term for the dilaton $X$ in (\ref{actionKEdil})
above. The Weyl transformation factors cancel in the derivative terms
$(\del\Phi)^2=g^{\mu\nu}\del_\mu\Phi\del_\nu\Phi$ with $\sqrt{-g}$
and the factor in $V$ gives $X^{-2{D-3\over D-2}+2}$\,.
The resulting action is in agreement with \eg\
\cite{Grumiller:2000wt,Grumiller:2001ea,Grumiller:2002nm} on the
s-wave reduction of Einstein gravity (dropping a term from the sphere
curvature) and others.

Thus finally we obtain the action
\be\label{actionXPhiU}
S= {1\over 16\pi G_2} \int d^2x\sqrt{-g}\, \Big(X^2\mathcal{R}
-\frac{1}{2} X^2 (\partial\Phi)^2-U(X,\Phi) \Big)\ ,
\ee
with $U=VX^{2/(D-2)}$ for the reduction of (\ref{actionDdim}) above.\\
More generally, $U(X,\Phi)$ is some general dilaton potential coupling
the dilaton $X$ to the scalar $\Phi$.\ The equations of motion from
this general action are
\bea\label{2dimseom-EMD0}
g_{\mu\nu}\nabla^2X^2-\nabla_{\mu}\nabla_{\nu}X^2
  +\frac{g_{\mu\nu}}{2}\Big(\frac{X^2}{2}(\partial\Phi)^2+U\Big)
  -\frac{X^2}{2}\partial_{\mu}\Phi\partial_{\nu}\Phi &=& 0\ ,\nonumber\\ [1mm]
\mathcal{R}-\frac{1}{2}(\partial\Phi)^2 -
\frac{\partial U}{\partial(X^2)} &=& 0\ ,\nonumber\\ [1mm]
\frac{1}{\sqrt{-g}}\partial_{\mu}(\sqrt{-g}\,X^2 \partial^{\mu}\Phi)
  -\frac{\partial U}{\partial\Phi} &=& 0\ ,
\eea
Without loss of generality, we can employ conformal gauge and take the
2-dim metric as
\be
ds^2 = g_{\mu\nu}dx^\mu dx^\nu = e^{f(t,r)} (-dt^2+dr^2)\ .
\ee
Then\ \
$\Gamma^t_{tt} = \Gamma^t_{rr} = \Gamma^r_{tr} = {1\over 2} {\dot f} ,\ \
\Gamma^t_{tr} = \Gamma^r_{rr} = \Gamma^r_{tt} = {1\over 2} f'$\ and\ 
${\cal R} = e^{-f(t,r)} \big( {\ddot f} - f'' \big)$.\\ Combining the
various components of the Einstein equations and simplifying gives
\bea\label{2dimseom-EMD-1}
(tr)&& \qquad  \del_t\del_rX^2 
- {1\over 2} f'\del_tX^2 - {1\over 2} {\dot f} \del_rX^2
+ {X^2\over 2}{\dot\Phi} \Phi' = 0\ , \nn\\
(rr+tt) && \ \ \
- \del_t^2X^2 - \del_r^2X^2 + {\dot f} \del_tX^2 + f' \del_rX^2
- {X^2\over 2} ({\dot\Phi})^2 - {X^2\over 2} (\Phi')^2 = 0 ,\qquad \nn\\ [1mm]
(rr-tt) && \ \ \
- \del_t^2X^2 + \del_r^2X^2 + e^f U = 0\ ,\\ [1mm]
(X)&& \quad\  \big( {\ddot f} - f'' \big)
- {1\over 2} (-({\dot\Phi})^2+(\Phi')^2)
- e^f \frac{\partial U}{\partial(X^2)} = 0 ,\qquad \nn \\ [1mm]
(\Phi)&& \quad\   - \del_{t}( X^2 \del_t\Phi) + \del_{r}( X^2 \del_r\Phi)
- e^f {\del U\over\del\Phi} = 0\ . \nn  
\eea
Here we have combined the $tt$- and $rr$-components of the Einstein
equations in (\ref{2dimseom-EMD0}): this form turns out to be
instructive and useful in what follows.

We expect that in the asymptotic regions, the backgrounds become
``nearly'' static: the fields are slowly varying. So turning the time
derivatives off in these regions gives
\bea\label{bndryConds}
- \del_r^2X^2 + f' \del_rX^2 - {X^2\over 2} (\Phi')^2 = 0 ,\ \
&&\ \ \del_r^2X^2 + e^f U = 0\ , \nn\\
- f'' - {1\over 2} (\Phi')^2 - e^f \frac{\partial U}{\partial(X^2)} = 0 ,
\ \ &&\ \  \del_{r}( X^2 \del_r\Phi) - e^f {\del U\over\del\Phi} = 0\ .
\eea
These equations define the asymptotic regions describing the background:
the $r$-behaviour is subject to these.\\
Now let us assume some Big-Crunch singularity arises: near such a
singularity, there is rapid time variation, approaching a divergence.
Thus taking the time derivative terms to be dominant (dropping all the
other terms) gives the near singularity behaviour described by
\be\label{nearSing-EMD-1}
- \del_t^2X^2 + {\dot f} \del_tX^2 - {X^2\over 2} ({\dot\Phi})^2 \sim 0 ,
\qquad  - \del_t^2X^2 \sim 0 ,\qquad 
 {\ddot f} + {1\over 2} ({\dot\Phi})^2 \sim 0 ,\qquad
- \del_{t}( X^2 \del_t\Phi) \sim 0 .\ \ 
\ee
Interestingly, this appears ``universal'': the dilaton potential $U$
governing the asymptotic behaviour of the background has disappeared.
Combining these equations gives
\be
- \del_t^2X^2 \sim 0 ,\qquad \del_t(X^2\del_tf)\sim 0\ ,\qquad
       {\ddot f} + {1\over 2} ({\dot\Phi})^2 \sim 0 ,
       \qquad \del_{t}( X^2 \del_t\Phi) \sim 0\ .
\ee
Taking $X^2\sim t^k$ gives $k(k-1)=0$.\ ($k=0$, consistent with\
$a=0$,\ $\al=0$, is time-independent)
Solving these shows that the cosmological singularity is governed by
a ``universal'' subsector:
\be\label{univSing}
X^2 \sim t ,\quad\ e^f\sim t^a ,\quad\ e^\Phi\sim t^\al ;\qquad 
a={\al^2\over 2}\ .
\ee
These can be thought of as a Big-Bang ($t>0$) or -Crunch ($t<0$): we
will often find it convenient to refer to a Crunch, but will
continue to use $t$, not bothering about its sign.

This universal subsector is like a cosmological attractor: the
behaviour near the singularity, when it exists, has this universal
scaling behaviour. The question of its existence can only be answered
by a detailed analysis of the above equations in full.
This full analysis will determine the precise values of the exponents
$a,\al$, and in particular the $tr$-equation which couples the $t$-
and $r$-sectors will answer the question of existence.
We will carry out this detailed analysis later (sec.4).

\medskip

\noindent {\bf \emph{Flat space Kasner cosmological singularities:}}\ \ 
The universal subsector (\ref{univSing}) above is in fact the reduction
to 2-dimensions of the well-known Kasner cosmological singularities
restricted to the ``mostly'' isotropic subcase, as we will see now.
These arise in Einstein gravity coupled to a scalar $\Phi$, with action
(\ref{actionDdim}) and no potential $V=0$. Reduction gives the 2-dim
action (\ref{actionXPhiU}) with $U=0$.
The Kasner exponents $p_i$ are related to $a,\al$ above.

The ``mostly isotropic'' Kasner cosmology in $D$-dimensions sourced by
a scalar $\Phi$ is
\bea\label{FlatKas}
&&  ds_D^2 = -dt^2 + t^{2p_1}dx_1^2 + t^{2p_2} \sum_{i=2}^{D-1} dx_i^2\ ,\qquad
e^\Phi=t^\al\ ,\nonumber\\
&&  p_1+(D-2)p_2=1 ,\qquad p_1^2+(D-2)p_2^2=1-{1\over 2}\al^2\ .
\eea
This is a 1-parameter family of Big-Crunch singularities, parametrized
by the scalar exponent $\al$.\ When $\al=0$, these have\
$p_1=-{1\over 3}\,, p_2={2\over 3}$ when $D=4$: this is the behaviour
near the spacelike curvature singularity in the interior of the
Schwarzschild black hole in 4-dim. See \cite{Frenkel:2020ysx} for a
recent discussion.

Upon reduction on $T^{D-2}\equiv \{ x_2,\ldots,x_{D-1}\}$,\ this gives,
using (\ref{ads2Redux1}), (\ref{weyl}),
\be
X^{{4\over D-2}}=t^{2p_2}\ ,\qquad
ds^2 = X^{{2(D-3)\over D-2}} \left(-dt^2 + t^{2p_1}dx_1^2\right)
= t^{2p_1+(D-3)p_2} \left( -dt^2/t^{2p_1}+dx_1^2 \right)\ .
\ee
Redefining gives
\be
T=t^{1-p_1}:\ \  X^2=T ,\qquad
ds^2=T^{((D-3)p_2+2p_1)/(1-p_i)} (-dT^2+dx_1^2) ,\qquad
e^\Phi=T^{\al/(1-p_1)}\,.
\ee
It can then be checked using (\ref{FlatKas}) that these exponents
satisfy (\ref{univSing}). The parametrization (\ref{univSing}) of
(\ref{FlatKas}) is just more convenient for this mostly isotropic
subclass of singularities it describes.

These backgrounds do not have any spatial dependence: so the full
equations (\ref{2dimseom-EMD-1}) in fact reduce to (\ref{nearSing-EMD-1}),
the asymptotic region equations (\ref{bndryConds}) being trivial.

\bigskip

\noindent {\bf \emph{$AdS$-Kasner singularities and redux:}}\\ 
The higher dimensional $AdS$ theories we have been discussing are
deformations of $AdS\times S$. The bulk theories are solutions to
Einstein gravity with the string dilaton $\Phi$, with action
\bea
S &=& {1\over 16 \pi G_D} \int d^Dx \sqrt{-g^{(D)}}\,
\big( {\cal R} - 2\Lambda - {1\over 2} (\del\Phi)^2 \big) \nonumber\\
&\longrightarrow&
    {1\over 16\pi G_2} \int d^2x \sqrt{-g} \left(X^2 {\cal R}
    - 2\Lambda X^{2/(D-2)} - {1\over 2} X^2 (\del\Phi)^2 \right)
\eea
using the reduction (\ref{ads2Redux1}), (\ref{weyl}).
Setting the $AdS$ scale in sec.~2 to unity $R_{AdS}=1$ makes $\Lambda$
also dimensionless: so this $AdS_D$ redux gives the 2-dim action
(\ref{actionXPhiU}) with
\be\label{potAdS}
U=2\Lambda X^{2/d_i}\ ,\qquad \Lambda=-{1\over 2}\,d_i(d_i+1)\ ,\qquad
D=d_i+2\ ,
\ee
Here $d_i$ is the spatial
dimension of the boundary theory, and we have
\bea\label{AdSDKas}
&& ds^2 = {1\over r^2} (-dt^2 + dr^2) + {t^{2p}\over r^2} \sum_i dx_i^2\ ,
\qquad e^\Phi=t^\al\ ,\nonumber\\
&& p={1\over d_i}\ , \qquad
\al = \sqrt{2(1-d_ip^2)} = \sqrt{{2(d_i-1)\over d_i}}\ .
\eea
The exponent $p$ is related to the dimension for these isotropic
$AdS$-Kasner backgrounds using (\ref{adsKasner}).\
This gives\
\be
ds^2_{AdS_{d_i+2}}={1\over r^2} (-dt^2+dr^2)+{t^{2/d_i}\over r^2} dx_i^2\ ,
\ee
so using (\ref{ads2Redux1}), (\ref{weyl}), gives the 2-dim background,
\be\label{AdSKas-XfPhi}
X^2 \sim {t\over r^{d_i}}\ ,\quad\
ds^2 = X^{{2(d_i-1)\over d_i}} ds_2^2
= {t^{(d_i-1)/d_i}\over r^{d_i+1}} (-dt^2+dr^2)\ ,\quad\
e^\Phi = t^{\sqrt{{2(d_i-1)/d_i}}}\ .
\ee
We see that the behaviour near the singularity $t\sim 0$ is precisely
as in (\ref{univSing}).

The precise value of the exponents $a,\al$ is fixed from the asymptotics,
using (\ref{bndryConds}): for instance, for the time-dependence to cancel
in $\del_r^2X^2+e^fU=0$, we require
\be
t^k \sim t^a t^{k/d_i} \quad\Rightarrow\quad a + {k\over d_i} = k\ \
\Rightarrow\ \ a = {d_i-1\over d_i}\ ,
\ee
using $k=1$.\
We will recover this and other solutions from a detailed analysis in
what follows.

\section{Solving the 2-dim theory for Big-Crunches}

We would now like to carry out a detailed analysis of backgrounds
with cosmological, Big-Crunch type, singularities, for theories of
the form (\ref{actionXPhiU}): we will discuss various kinds of
potentials $U$ that arise from familiar theories under KK-reduction.\\
Towards this, we take the following scaling-type ansatz
\begin{equation}\label{XfPhi-ansatz}
  X^2=t^kr^m \ ,\qquad e^f=t^ar^b\ ,\qquad e^{\Phi}=t^{\alpha}r^{\beta}\ .
\end{equation}
In doing this, we have set relevant lengthscales to unity.
For instance, we have set $R_{AdS}=1$, recasting (\ref{adsKasner}) as
(\ref{AdSDKas}): then the metric is dimensionless.
It is straightforward to reinstate the lengthscales when necessary.\

For time-independent cases, this is in accord with the kinds of power
law behaviour in known backgrounds such as $AdS$ or conformally $AdS$,
and also consistent with the asymptotic (time-independent) form of the
equations (\ref{bndryConds}).
For time-dependent situations, our intuition is that the power law
$t$-behaviour implies a Crunch (or Bang) at some time $t=0$: so in
sense, the ansatz (\ref{XfPhi-ansatz}) is a ``near singularity''
ansatz describing the behaviour near the cosmological singularity,
consistent with the universal near singularity form (\ref{nearSing-EMD-1}),
(\ref{univSing}) of the equations of motion (\ref{2dimseom-EMD0}),
(\ref{2dimseom-EMD-1}). Juxtaposing the time part is consistent with
the apparent separability of the equations into time and space parts,
and we will in fact find that this is consistent. However this does
not of course include the possibility of more exotic time-dependent
backgrounds which do not admit any factorization.

The scaling ansatz (\ref{XfPhi-ansatz}) gives
\begin{equation}
  \dot{f}=\frac{a}{t} \ , \quad f^{\prime}=\frac{b}{r}\qquad
              {\rm and}\qquad
  \dot{\Phi}=\frac{\alpha}{t} \ ,\quad \Phi^{\prime}=\frac{\beta}{r}\ .
\end{equation}
Using this and putting the scaling ansatze (\ref{XfPhi-ansatz}) back
in the differential equations \eqref{2dimseom-EMD-1} and requiring
that nontrivial solutions exist for all $t, r$, gives a set of
\emph{algebraic} equations for the various exponents.  Solving for the
exponents then determines the full solution.\ 
In a sense, this is akin to the Frobenius series solutions for
differential equations, with the additional assumption that it is
adequate to focus on the leading monomials in a series \eg\
$\sum_i t^{m_i}r^{n_i}$.\ This is consistent for both the asymptotic
region (large $r$) and the near singularity region ($t\ra 0$) where
appropriate leading terms can be taken to be good approximations.

In practice, for time-dependent backgrounds, these equations can be
solved by first taking the $t$- and $r$-sectors as decoupled: this
yields solutions for the $t$- and $r$-exponents, which can then be
stuck into the $tr$-equation which is the only equation coupling the
$t$- and $r$-sectors. We will see this explicitly below.

\subsection{$AdS$ and conformally $AdS$ backgrounds}

Consider the class of models given by the higher dimensional action
\bea\label{NonconfDp}
&&   S = \int d^{d_i+2}x\sqrt{-g}\,\Big({\cal R} - {1\over 2}(\del\Phi)^2
- 2\Lambda e^{\gamma\Phi}\Big) ,\qquad
\Lambda = -{1\over 2}(d_i+1-\theta)(d_i-\theta) ,\nn\\
&& 
\quad \gamma = {-2\theta\over\sqrt{2d_i(d_i-\theta)(-\theta)}}\ ,
\qquad d_i=p=D-2\,,\qquad \theta=p-{9-p\over 5-p}<0\ .
\eea
For $\theta=0$ these describe conformal branes: $\gamma=0$ and the
potential in $\Phi$ is simply the negative cosmological constant term
arising from the flux integrated over the transverse sphere.
The parameter $\theta$ is called the ``hyperscaling violation'' exponent:
these theories are a subclass of more general hyperscaling violating
Lifshitz theories \cite{Taylor:2015glc,Hartnoll:2016apf} which we will
discuss later. For nonzero $\theta$, these arise from nonconformal
$Dp$-branes \cite{Itzhaki:1998dd} after dimensional reduction over the
transverse sphere $S^{8-p}$\ \cite{Dong:2012se}.\
The known time-independent backgrounds are $(p+2)$-dimensional spaces
of the form
\be\label{confAdS}
ds^2 = r^{2\theta/d_i} \Big({-dt^2 + dx_i^2+dr^2\over r^2}\Big) ,
\qquad e^\Phi = r^{\sqrt{2(d_i-\theta)(-\theta/d_i)}}\ .
\ee
For $\theta=0$ these give $AdS_{p+2}$ with $\Phi=const$. Nonzero $\theta$
gives an extra conformal factor and a nontrivial string dilaton $\Phi$
which is required to source such backgrounds: it is also related to the
RG running of the dual field theory. Besides $R_p$ akin to $R_{AdS}$
earlier, there is a new length scale here $r_{hv}$ which appears in all
$\theta$-dependent terms: \eg\ the conformal factor is
$({r\over r_{hv}})^{2\theta/d_i}$, and likewise for $e^\Phi$. It is
straightforward to reinstate $R_p$ and $r_{hv}$.


The action above is of the form (\ref{actionDdim}): under reduction
with (\ref{ads2Redux1}), (\ref{weyl}), we obtain a 2-dimensional
dilaton-gravity-matter theory of the form (\ref{actionXPhiU}). The
dilaton potential here is
\be\label{pot-c1}
U(X,\Phi) = -c_1X^{\frac{2}{d_i}}e^{\gamma\Phi}\ , \qquad
c_1=-2\Lambda=(d_i+1-\theta)(d_i-\theta)\ .
\ee
For $\gamma=0$, we see that this becomes (\ref{potAdS}) as expected.\
We have
\be
\frac{\partial U}{\partial(X^2)} = -\frac{c_1}{d_i}(X^2)^{\frac{1}{d_i}-1}e^{\gamma\Phi}\ , \qquad
\frac{\partial U}{\partial\Phi} = -\gamma c_1X^{\frac{2}{d_i}}e^{\gamma\Phi}\ .
\ee
Now using the ansatz (\ref{XfPhi-ansatz}) in (\ref{2dimseom-EMD-1}) gives
the algebraic equations:
\bea\label{EOM-c1-trExp}
(tr) && \left[mk-\frac{bk}{2}-\frac{am}{2}+\frac{\alpha\beta}{2}\right] t^{k-1}r^{m-1}=0\ , \nn\\
(rr+tt) && \left[ -k(k-1)+ak-\frac{\alpha^2}{2}\right] t^{k-2}r^m +
\left( -m(m-1)+bm-\frac{\beta^2}{2}\right) r^{m-2}t^k=0\ ,\qquad \nn\\
(rr-tt) && \big[-k(k-1)\big]\frac{r^2}{t^2} + m(m-1) - c_1t^{a+k(1/d_i-1)+\gamma\alpha}r^{b+m(1/d_i-1)+\gamma\beta + 2}=0\ , \\
(X) && -\frac{r^2}{t^2}\left[a-\frac{\alpha^2}{2}\right] +
b-\frac{\beta^2}{2} +
\frac{c_1}{d_i}t^{a+k(1/d_i-1)+\gamma\alpha}r^{b+m(1/d_i-1)+\gamma\beta + 2}=0\ , \nn\\
(\Phi) && \big[-\alpha(k-1)\big]\frac{r^2}{t^2} + \beta(m-1)
+ \gamma c_1t^{a+k(1/d_i-1)+\gamma\alpha}r^{b+m(1/d_i-1)+\gamma\beta + 2}=0\ .\nn
\eea
In the above equations, we have demarcated the time-dependent terms in
square brackets for convenience. Then for time-independent solutions,
all the square bracket terms above must vanish: in particular the
$(tr)$-equation is automatically satisfied while the other equations
give nontrivial constraints on the $r$-behaviour.


As stated earlier, for time-dependent backgrounds, we first take the
$t$- and $r$-sectors as decoupled, thereby obtaining solutions for the
$t$- and $r$-exponents. These can then be stuck into the $tr$-equation
which is the only equation coupling the $t$- and $r$-sectors.

Thus, from the $r$- and $t$-equations above, requiring that the various
$t^Ar^B$ terms either cancel in groups or vanish (for non-matching
exponents), we find the following nontrivial family of solutions,
characterized by relations between the exponents $(k,a,\al)$ and
$(m,b,\beta)$ and the parameters $d_i,\theta,c_1$ of the theory:
\bea\label{bbetamaal-1}
(r) && b=\frac{\beta^2}{2}-\frac{m(m-1)}{d_i} \,; \qquad
    bm=\frac{\beta^2}{2}+m(m-1) \,;
  \qquad b+\frac{m(1-d_i)}{d_i}+\gamma\beta=-2\,;\nn \\ [1mm]
  &&\qquad\qquad\qquad\qquad
  (\beta + \gamma m) (m-1) = 0 \,;\qquad m(m-1)=c_1 ;\\ [3mm]
(t) &&  k(k-1)=0\ ;\qquad a=\frac{\alpha^2}{2}\ ;\qquad
  \al(k-1)=0\ ;\qquad  a+\frac{k(1-d_i)}{d_i}+\gamma\alpha=0 .\
  \label{bbetamaal-2}
\eea
The $(t)$-equations arise from the square bracket terms in
(\ref{EOM-c1-trExp}). The last equation here arises from noting that
the exponent of $t$ in the terms containing $c_1$ must independently
vanish for a nontrivial solution to exist for all $t, r$\ (thus
matching the vanishing $t$-exponent in the other terms), as stated
above.\\
Likewise the $(r)$-equations arise from noting that the terms outside
the square brackets must vanish for generic $t^Ar^B$ terms: in
particular this also requires that the exponent of $r$ in the terms
containing $c_1$ must vanish (as for $t$ above), leading to the first
two $(r)$-equations.\ Finally we have used the equation $m(m-1)=c_1$
from the $(rr-tt)$-equation in (\ref{EOM-c1-trExp}) to recast the
above equations (\ref{bbetamaal-1}), (\ref{bbetamaal-2}), in a
suggestive and useful manner.

\medskip

\noindent \underline{Time-independent backgrounds:}\ \
We will first discuss time-independent solutions, recovering the familiar
$AdS$ and nonconformal $Dp$-brane backgrounds. This also serves as a
check of these equations and our scheme of analysing them.\
In this case, we have $k=0,\ a=0,\ \al=0$: the last equation in
(\ref{bbetamaal-2}) is in fact not applicable since there is no
$t$-exponent in the absence of time-dependence.
These imply that the $tr$-equation is automatically satisfied.

Massaging the set of relations between $b,\beta,\gamma$ in
(\ref{bbetamaal-1}) having eliminated $c_1$ as stated above shows that
the relations are not all independent, allowing for consistent
nontrivial solutions. The first two equations in (\ref{bbetamaal-1})
parametrize $b,\beta$ in terms of $m$: then using $\gamma$ from
(\ref{NonconfDp}) gives
\be\label{b,beta}
b={m(1+d_i)\over d_i}\ ,\qquad
\beta=-m\gamma = \sqrt{{2m(m+d_i)\over d_i}}\ ,\qquad m=-d_i+\theta\ .
\ee
Note $m<0$, since $\theta\leq d_i$ from the entropy-temperature
relation\ $S\sim V_{d_i}T^{d_i-\theta}$ and specific heat positivity
\cite{Dong:2012se}.
Note also that $m$ obtained from $m(m-1)=c_1$ (using (\ref{pot-c1})) gives
\be\label{mc1}
m^2-m-(d_i+1-\theta)(d_i-\theta)=0\ \ \ \Rightarrow\ \ \
m=-(d_i-\theta)\ \ {\rm or}\ \ (d_i-\theta+1)\ .
\ee
The first $m$ value corresponds to non-normalizable modes in the higher
dimensional theory while $m=d_i-\theta+1$ are normalizable modes.
Of these, we see that only the non-normalizable $m$ value satisfies
all the equations in (\ref{bbetamaal-1}), (\ref{bbetamaal-2}). Note
also that the possible value $m=1$ satisfies all equations except for
$c_1=m(m-1)$ with nonzero $c_1$.

These in fact recover the reduction of the backgrounds in (\ref{confAdS}),
using (\ref{ads2Redux1}), (\ref{weyl}) and (\ref{XfPhi-ansatz}). For
instance, $X^{4/d_i}=r^{2\theta/d_i-2}$ gives $m=-d_i+\theta$, while
$\beta=\sqrt{2(d_i-\theta)(-\theta/d_i)}$ matches with the $e^\Phi$
exponent. Likewise $b$ matches with the $e^f$ exponent after noting
that the piece ${m(1-d_i)\over d_i}$
in (\ref{bbetamaal-1}) arises from the Weyl transformation (\ref{weyl})\
(likewise for the ${k(1-d_i)\over d_i}$ in (\ref{bbetamaal-2})).

\medskip

Now coming to cosmological solutions with time-dependence, we see the
universal sector $k=1,\ a={\al^2\over 2}$ from the first two
$(t)$-equations, while the last equation fixes the precise value
of $a, \al$.
These show a Big-Crunch type cosmological singularity at $t=0$ with
$X^2\sim t$ always. We will now discuss various subcases here.


\medskip

\noindent  \underline{$AdS_D$ Big-Crunch redux:}\ \ Here $\theta=0$
and so $\gamma=0=\beta$. Then (\ref{b,beta}) and the time-dependent
subsector (\ref{bbetamaal-2}) can be seen to give the 2-dim background 
(\ref{XfPhi-ansatz}) with the exponents
\be\label{AdSD-exp1}
\theta=0=\gamma, \qquad \beta=0, \qquad m=-d_i ,\qquad b=-(d_i+1)\ ,
\ee
\be\label{AdSD-exp2}
k=1 ,\qquad a={\al^2\over 2}={d_i-1\over d_i}\ .
\ee
The $tr$-equation\ $m-{b\over 2}-{am\over 2}-{\al\beta\over 2}=0$ is
explicitly seen to be satisfied.
Thus we recover the $AdS$ Kasner cosmological singularities
(\ref{AdSDKas}) in the form (\ref{AdSKas-XfPhi}) after reduction
to 2-dimensions, noting $D=d_i+2$. We have chosen $\al>0$ so $e^\Phi\ra 0$
near the singularity.

\medskip

\noindent \underline{Nonconformal $Dp$-brane Big-Crunch redux:}\ \
Now we have $\theta\neq 0$, the higher dimensional theories being
(\ref{NonconfDp}). The time-sector (\ref{bbetamaal-2}) now contains
$\gamma$ as well, and can be solved for $a, \al$.
The $r$-sector, decoupled from the $t$-sector, can be solved as
for the time-independent case in (\ref{b,beta}). Putting these together
gives the 2-dim background (\ref{XfPhi-ansatz}) with exponents
\bea\label{Dp-mbbetakaal}
&& m=-(d_i-\theta)\ ,\qquad b=-{(d_i-\theta)(1+d_i)\over d_i}\ ,\qquad
\beta=-m\gamma=\sqrt{{2(d_i-\theta)(-\theta)\over d_i}}\ ,\nn\\
&& \qquad k=1 ,\qquad a={\al^2\over 2}\ ,\qquad
\al=-\gamma \pm \sqrt{\gamma^2+{2(d_i-1)\over d_i}}\ .
\eea
(recall $\theta<0$, $\gamma>0$, from (\ref{NonconfDp}))\
The $tr$-equation, using (\ref{bbetamaal-1}), (\ref{bbetamaal-2}), then
gives
\be\label{tr-gamma}
m - {b\over 2} - {am\over 2} + {\al\beta\over 2} =
m + {m\over 2} \Big({1-d_i\over d_i}+\gamma\al \Big)
- {1\over 2} \Big(-2-m{1-d_i\over d_i} + m\gamma^2 \Big) - {m\gamma\al\over 2}
\ee
We see that $\al$ drops out of this entirely: the remaining terms can
be seen to vanish, using the exponents (\ref{Dp-mbbetakaal}) above: thus
the $tr$-equation is satisfied here as well.

Now using (\ref{ads2Redux1}) and undoing the Weyl transformation
(\ref{weyl}) gives the higher dimensional theory corresponding to
the 2-dim Big-Crunch (\ref{XfPhi-ansatz}) with exponents above: we obtain
\be\label{Nconf-t}
ds^2 = {e^f\over X^{{2(d_i-1)\over d_i}}}(-dt^2+dr^2)+X^{{4\over d_i}}dx_i^2 = 
{r^{{2\theta\over d_i}}\over r^2} \Big( {-dt^2+dr^2\over t^{\gamma\al}} +
  t^{{2\over d_i}} dx_i^2\Big) ,\quad
e^\Phi=t^\al r^{\sqrt{2(d_i-\theta)({-\theta\over d_i})}}\,,
\ee
noting that the $r$-exponents are the same as in the time-independent
backgrounds stated earlier. Note that the $tr$-part of the spacetime
has $t$-behaviour opposite to that of $e^\Phi$ since $\gamma>0$: thus
taking the positive sign in $\al$ above so that $\al>0$ leads to
$e^\Phi\ra 0$ as $t\ra 0$ but the $tr$-part expands (although the
$x_i$-part crunches). Conventionally this is what would be called
a Big-Crunch (as in the $AdS$ case earlier).

\subsection{Hyperscaling violating Lifshitz asymptotics}

We now consider another class of nonrelativistic theories, hyperscaling
violating Lifshitz (hvLif) ones. Consider the action
\bea\label{EMD-gl1}
&& \qquad S = \int d^{d_i+2}x\sqrt{-g}\,\Big({\cal R} - {1\over 2}(\del\Phi)^2
- {1\over 4} Z(\Phi) F^2 - 2\Lambda e^{\gamma\Phi} \Big)\ ,
\nonumber\\
&&\qquad\qquad\qquad Z(\Phi) = e^{-\lambda\Phi}\ ,\qquad\qquad
\Lambda=-{1\over 2} (d_i+z-\theta)(d_i+z-\theta-1) , \nonumber\\ 
&& \gamma = {-2\theta\over\sqrt{2d_i(d_i-\theta)(d_iz-d_i-\theta)}}\ ,
\qquad\quad  \lambda = {-2(\theta + d_i(d_i-\theta))\over
  \sqrt{2d_i(d_i-\theta)(d_iz-d_i-\theta)}}\ .\qquad\quad
\eea
These are Einstein-Maxwell-Dilaton theories, containing gravity
(and $\Lambda<0$) sourced by a gauge field $F$ and a scalar $\Phi$.
Such theories
have been actively studied in the last several years in the context
of ``AdS/CMT'' towards understanding nonrelativistic generalizations
of holography and condensed-matter-type applications. See \eg\
\cite{Taylor:2015glc,Hartnoll:2016apf} for a review of some of these
developments\ (our conventions for the higher dimensional theories in
this section are as in \cite{Mukherjee:2017ynv}). These admit hvLif
spacetimes,
\be\label{hvLbgnd}
ds^2 = \rho^{2\theta/d_i} \Big(-{dt^2\over \rho^{2z}}
+ {dx_i^2+d\rho^2\over \rho^2}\Big) ,
\ \ \
A_{1,t} = -\sqrt{{2(z-1)\over d_i+z-\theta}}\,{1\over \rho^{d_i+z-\theta}}\,,
\ \ \ e^\Phi = \rho^{\sqrt{2(d_i-\theta)(z-\theta/d_i-1)}}\,,
\ee
where $z$ is the Lifshitz exponent and $\theta$ the hyperscaling
violating one.
In the subsector with the gauge field background profiles taken on-shell,
these theories give theories with effective action of the form
(\ref{actionDdim}): this is similar in spirit to the 5-form field strength
giving rise to an effective cosmological constant in 5-dimensions for
a D3-brane stack. This is described in greater detail in
\cite{Kolekar:2018sba,Kolekar:2018chf}.
Dimensional reduction of these effective theories using (\ref{ads2Redux1}),
(\ref{weyl}), leads to 2-dimensional dilaton-gravity-matter theories of
the form (\ref{actionXPhiU}) with the potential taking the form
\be\label{pot-c1c2}
U =  X^{2/d_i} \Big(-c_1 e^{\gamma\Phi} + {c_2\over X^4} e^{\lambda\Phi}\Big)\ .
\ee
For $z=1$, these are conformally $AdS$ spaces of the form (\ref{confAdS})
discussed earlier: in this case, $c_2$ vanishes.

With the potential (\ref{pot-c1c2}), we obtain the following equations
from (\ref{2dimseom-EMD-1}) using the scaling ansatz (\ref{XfPhi-ansatz}):
\bea\label{t&r-c1+c2}
(tr) && \left[ mk-\frac{bk}{2}-\frac{am}{2}+\frac{\alpha\beta}{2}\right]
t^{k-1}r^{m-1}=0\ , \nn\\
(rr+tt) && \left[ -k(k-1)+ak-\frac{\alpha^2}{2}\right] t^{k-2}r^m
+ \left( -m(m-1)+bm-\frac{\beta^2}{2}\right) r^{m-2}t^k=0\ , \qquad \nn\\
\nonumber (rr-tt) && \big[-k(k-1)\big] \frac{r^2}{t^2} + m(m-1)
- c_1\,t^{a+k(1/d_i-1)+\gamma\alpha}r^{b+m(1/d_i-1)+\gamma\beta + 2} \nn\\
&& \hspace{4.5cm} +\, c_2\, t^{a+k(1/d_i-3)+\lambda\alpha}
\times  r^{b+m(1/d_i-3)+\lambda\beta + 2}=0\ ,\quad \\
\nonumber (X) && -\frac{r^2}{t^2}\left[a-\frac{\alpha^2}{2}\right]
+ b-\frac{\beta^2}{2} + \frac{c_1}{d_i}\, t^{a+k(1/d_i-1)+\gamma\alpha}r^{b+m(1/d_i-1)+\gamma\beta + 2} \nn\\
&& \hspace{4cm} - c_2\Big({1\over d_i}-2\Big)t^{a+k(1/d_i-3)+\lambda\alpha}r^{b+m(1/d_i-3)+\lambda\beta + 2}=0\ , \nn\\
(\Phi) && \big[-\alpha(k-1)\big]\frac{r^2}{t^2} + \beta(m-1)
+ \gamma c_1\, t^{a+k(1/d_i-1)+\gamma\alpha}r^{b+m(1/d_i-1)+\gamma\beta + 2} \nn\\
&& \hspace{5cm} - \lambda c_2\, t^{a+k(1/d_i-3)+\lambda\alpha}r^{b+m(1/d_i-3)+\lambda\beta + 2}=0\ . \nn
\eea
For $c_2=0$, these reduce to (\ref{EOM-c1-trExp}). As in that case, we
have demarcated the time-dependent terms in square brackets.
These and the other equations give the following algebraic equations
between the various exponents $(k,a,\al)$ and $(m,b,\beta)$ and the
parameters $c_1,c_2,\gamma,\lambda$ of these theories:
\begin{eqnarray}
(tr) && \hspace{5cm}
mk-\frac{bk}{2}-\frac{am}{2}+\frac{\alpha\beta}{2} = 0\ ,\label{tr-c1c2}\\ [2mm]
(r) &&\ \
bm = \frac{\beta^2}{2} + m(m-1) ,\quad\
b + m\Big({1-d_i\over d_i}\Big) + \gamma\beta + 2 = 0 ,\quad\
(\lambda - \gamma)\beta = 2m ,\qquad \label{bbetam}\\ [1mm]
&& \ \
b-\frac{\beta^2}{2} - m(m-1)\Big({1\over d_i}-2\Big) - 2c_1 = 0 , \quad
\beta (m-1) + \gamma c_1 - \lambda c_2 = 0 ,\qquad
\label{bbetam2}\\ [1mm]
&& \hspace{6cm} c_2 = c_1 - m(m-1) , \label{c2c1m}\\ [2mm]
(t) && -k(k-1)+ak-\frac{\alpha^2}{2} = 0 ,\qquad k(k-1) = 0 , \qquad
a = \frac{\alpha^2}{2}\,,\qquad \alpha(k-1) = 0 , \nn\\
&& \qquad\qquad a + \Big({1\over d_i}-1\Big) + \gamma\alpha = 0\,,\quad
a+\Big({1\over d_i}-3\Big)+\lambda\alpha = 0\,.\label{t-c1c2}
\end{eqnarray}
These are obtained, as before, by requiring the exponents of various
$t^Pr^Q$-terms to be equal for a nontrivial solution valid for all
$t, r$, implying that various $t$-exponents independently vanish.\
Note that, compared with (\ref{EOM-c1-trExp}), (\ref{bbetamaal-1}),
(\ref{bbetamaal-2}), we obtain more constraining equations in this
case, stemming from the $c_2$ term. In the above, the last equation
in (\ref{bbetam}) is obtained by combining the second one there with
a similar equation $b + m({1-3d_i\over d_i}) + \lambda\beta + 2 = 0$
arising from the $r$-exponent in the $c_2$-term in the $(rr-tt)$-,\
$(X)$- and $(\Phi)$-equations in (\ref{t&r-c1+c2}).\\
As in (\ref{bbetamaal-1}), (\ref{bbetamaal-2}), we have grouped the
$r$- and $t$-equations separately: the $(tr)$-equation couples them.

Restricting to the time-independent case, we ignore (\ref{tr-c1c2}),
(\ref{t-c1c2}), thereby obtaining the equations (\ref{bbetam}),
(\ref{bbetam2}), (\ref{c2c1m}).\
The 3 equations in (\ref{bbetam}) can be solved (and simplified in
\eg\ Mathematica), using the $\gamma, \lambda$-values in
(\ref{EMD-gl1}) to give
\be\label{hvL-mbbeta-r}
m = {-d_i+\theta\over z}\, ,\quad\ \
\beta = {\sqrt{2(d_i-\theta)(z-1-\theta/d_i)}\over z}\, ,\quad\ \
b = {\theta + d_i (1-d_i + \theta - 2z)\over d_i z}\, .
\ee
It can be checked that these also solve the equation obtained by
eliminating $c_2$ using (\ref{c2c1m}) and then $c_1$ from the two
equations in (\ref{bbetam2}) above. Using (\ref{hvL-mbbeta-r}),
$c_1, c_2$ are found as
\be
c_1 = {(d_i+z-\theta)(d_i+z-\theta-1)\over z^2} = -{2\Lambda\over z^2}\ ,
\qquad  c_2 = {(z-1)(d_i+z-\theta)\over z^2}\ ,
\ee
with $\Lambda$ in (\ref{EMD-gl1}).
To compare with the reduction (\ref{ads2Redux1}), (\ref{weyl}), of
hvLif backgrounds, write (\ref{hvLbgnd}) as
\be
ds^2 = \rho^{2\theta/d_i-2z} (-dt^2+\rho^{2z-2}d\rho^2)
+ \rho^{2\theta/d_i-2} dx_i^2\quad
\xrightarrow{r\sim \rho^z/z}\quad r^{\#} (-dt^2+dr^2) + r^{\#} dx_i^2\ .
\ee
These can be seen to match.

\medskip

\noindent \underline{Cosmological solutions:}\ \ These are more intricate
and constrained, since the $r$- and $t$-exponents are now apparently
coupled by the $(tr)$-equation which is more constraining in this case.
From the last equation in (\ref{bbetam}) and (\ref{t-c1c2}), we obtain
\be
(\lambda - \gamma)\beta = 2m\ ,\qquad (\lambda - \gamma)\alpha = 2\ 
\qquad \Rightarrow \quad \beta = m\alpha \ .
\ee
Then the first equation in (\ref{bbetam}) alongwith $a={\al^2\over 2}$
from (\ref{t-c1c2}) gives
\be
bm=m^2a+m(m-1) \quad \Rightarrow\quad b=m(a+1)-1\ ,
\ee
taking $m\neq 0$.\ 
Sticking these in the $(tr)$-equation (\ref{tr-c1c2}) then gives
\be
m - {m(a+1)-1\over 2} - {am\over 2} + ma = 0 \quad\Rightarrow\quad
m=-1\ .
\ee
\be\label{mbabetaal}
\Rightarrow\quad m=-1 ,\qquad \beta=-\alpha=-{2\over \lambda - \gamma}\,,
\qquad b=-a-2\ .
\ee
Then the second equation in (\ref{bbetam2}) alongwith (\ref{c2c1m}) gives
\be
c_1-c_2=2 ,\quad 2\al+\gamma c_1 -\lambda c_2=0\quad\Rightarrow\quad
c_1={2(\al+\lambda)\over \lambda-\gamma}\,,\quad
c_2={2(\al+\gamma)\over \lambda-\gamma}\ .
\ee
Using these in the first equation in (\ref{bbetam2}) alongwith
(\ref{mbabetaal}), after some simplification, gives
\be
\beta = \lambda \Big({1\over d_i}-1\Big) - \gamma \Big({1\over d_i}-3\Big)\ .
\ee
Note that consistency of this $\beta$ value with that in
(\ref{mbabetaal}) gives a nontrivial relation between the parameters
$\gamma,\lambda$ of these theories: see below.
Finally note that using $a={\al^2\over 2}$ in the two equations in the
second line in (\ref{t-c1c2}) gives two quadratics, with solutions
\be\label{hvL-t-al}
\al = -\gamma-\sqrt{\gamma^2+{2(d_i-1)\over d_i}}\ ,\qquad {\rm and}\qquad
\al = -\lambda-\sqrt{\lambda^2+{2(3d_i-1)\over d_i}}\ ,
\ee
which are consistent only if equal: this after some simplification
can be shown to give
\be\label{gammalambda1}
\lambda \Big({1\over d_i}-1\Big) - \gamma \Big({1\over d_i}-3\Big) 
= {-2\over \lambda - \gamma}\ .
\ee
This is the same nontrivial condition on $\gamma, \lambda$, as the
one from $\beta$ stated above.\
Taking this as a nontrivial quadratic relation in $\lambda$, we obtain
\be
\lambda = {(2d_i-1)\gamma - \sqrt{d_i^2(2+\gamma^2)-2d_i}\over d_i-1}\ .
\ee
Now finally, we will check consistency of these exponent values with
the parametrization of $\gamma, \lambda$ in (\ref{EMD-gl1}). We have\
$\lambda-\gamma={-2d_i(d_i-\theta)\over\sqrt{\ldots}}<0$ since
$d_i-\theta>0$ from the positivity of specific heat for hvLif theories
\cite{Dong:2012se}.
This requires $\al<0$ and $\beta>0$. This is why we have displayed only
the negative signs in the $\al$ values as consistent with (\ref{EMD-gl1}):
this further implies the negative sign in the $\lambda$ solution above.
Now
\be
\lambda-\gamma
= {d_i\over d_i-1} \left(\gamma - \sqrt{\gamma^2+2-{2\over d_i}}\right) < 0\ ,
\ee
which is consistent with the parametrization (\ref{EMD-gl1}).
Additionally, $m=-1$ implies $d_i-\theta=z$.

In particular for $\gamma=0$ and $d_i=2$ for simplicity, we have the
nontrivial Lifshitz-type cosmological solution (\ref{XfPhi-ansatz}) with
\be
\theta=0 ;\ \ \lambda=-2={-2d_i^2\over\sqrt{2d_i^3(z-1)}}\ \Rightarrow z=2\,;
\quad \beta=1=-\al ,\ b=-{5\over 2}\,,\ a={1\over 2}\,,\ k=1,\ m=-1\,.
\ee
Also $c_1=3,\ c_2=1\ ,\ \ \Lambda=-6$.\ It can be checked that these
are consistent with all the equations above.

In terms of $\rho=(2r)^{1/2}$, we have\ (using (\ref{ads2Redux1}) and
undoing the Weyl transformation (\ref{weyl}))
\bea
&&\qquad X^2={t\over\rho^2}\,,\ \ e^f={\sqrt{t}\over \rho^5}\,,\quad\
e^\Phi={\rho^2\over t} \nn\\
&& \Rightarrow\quad
ds^2 = {e^f\over X^{2(d_i-1)/d_i}}(-dt^2+dr^2)+X^{4/d_i}dx_i^2 = 
-{dt^2\over\rho^4} + {d\rho^2\over\rho^2} + {t\over\rho^2}dx_i^2\ .\qquad
\eea
In this case, since $\gamma=0$, the $t$-factors in $e^f$ and $X^\#$
cancel (stemming from (\ref{t-c1c2})): more generally there will be
$t$-dependence in the $tr$-part as well (as in (\ref{Nconf-t})). Then
the 2-dim cosmological background (\ref{XfPhi-ansatz}) has exponents
exhibiting the universal behaviour $k=1,\ a={\al^2\over 2}$ alongwith
(\ref{hvL-mbbeta-r}), (\ref{mbabetaal}), (\ref{hvL-t-al}) and 
(\ref{gammalambda1}) subject to the parametrization (\ref{EMD-gl1}).
This gives the higher dimensional cosmological hvLif spacetime
\be\label{hvLif-t}
ds^2 = {r^{2\theta/d_i}\over r^2} \left( {1\over t^{\gamma\al}}
\Big(-{dt^2\over r^{2z}}+dr^2\Big) + t^{2/d_i} dx_i^2\right) ,\qquad
e^\Phi=t^\al r^{-\al}\ ,
\ee
with $\al$ determined, and $z, \theta$ constrained as above.

\subsection{JT gravity}

It is instructive to study $AdS_2$ boundary conditions and look for
cosmological solutions: it is well known that extremal black holes or
branes give rise to $AdS_2$ in their throat regions. Performing a
dimensional reduction on the transverse space allows focussing on
these $AdS_2$ dilaton-gravity theories in 2-dimensions: this is the
Jackiw-Teitelboim gravity theory \cite{Jackiw:1984je,Teitelboim:1983ux}
and has been much studied lately in the context of SYK models and nearly
$AdS_2$ theories
\cite{Almheiri:2014cka,Maldacena:2016upp,Jensen:2016pah,Engelsoy:2016xyb}\
(see \eg\ \cite{Sarosi:2017ykf,Rosenhaus:2018dtp,Trunin:2020vwy} for
some reviews).

The JT gravity theory, with dilaton $X^2$, admits $AdS_2$ boundary
conditions: the action is
\be\label{2dPhi}
S= {1\over 16\pi G_2} \int d^2x\sqrt{-g}\, \Big[X^2 \big({\cal R} + 2\big)
  \Big]\ .
\ee
This is essentially the 2-dimensional action (\ref{actionXPhiU}) 
with dilaton potential\ $U=-2X^2$: this is of the form (\ref{pot-c1})
with $c_1=2,\ \gamma=0,\ d_i=2$.\ For simplicity, we are not including
the extra scalar $\Phi$\ (it turns out that a free 2-dim scalar $\Phi$
without the $X^2$ coupling in (\ref{actionXPhiU}) and additional
interaction decouples from the dynamics in this scaling analysis).

The equations of motion are similar to (\ref{2dimseom-EMD0}),
(\ref{2dimseom-EMD-1}), and give in this case:
\bea\label{2dPhi-eom}
(tr)&& \qquad  \del_t\del_rX^2 
- {1\over 2} f'\del_tX^2 - {1\over 2} {\dot f} \del_rX^2 = 0\ , \nn\\
(rr+tt) && \ \ \
- \del_t^2X^2 - \del_r^2X^2 + {\dot f} \del_tX^2 + f' \del_rX^2 = 0\ ,\qquad
\nn\\
(rr-tt) && \ \ \
- \del_t^2X^2 + \del_r^2X^2 + e^f U = 0\ ,\\
(X)&& \quad\  \big( {\ddot f} - f'' \big)
- e^f \frac{\partial U}{\partial(X^2)} = 0\ .\qquad \nn
\eea
These are the dilaton equation and the Einstein equations, combined.
As before, we look for Big-Crunch type behaviour, defined by large time
derivatives, approaching a divergence (thereby dropping the other terms).
Then, in contrast with (\ref{nearSing-EMD-1}), (\ref{univSing}), we see
the ``near singularity'' behaviour as
\be
- \del_t^2X^2 + {\dot f} \del_tX^2 \sim 0\ ,
\quad  - \del_t^2X^2 \sim 0\ ,\quad {\ddot f} \sim 0 \qquad\Rightarrow\qquad
f=at ,\quad X^2=t\ .
\ee
Instead of (\ref{XfPhi-ansatz}), this suggests the scaling ansatz
\be
X^2=t^kr^m ,\qquad e^f=e^{at}r^b .
\ee
Analysing as in the previous cases from redux, we stick in this
scaling ansatz in the equations (\ref{2dPhi-eom}) and obtain the
algebraic equations:
\bea
mk t^{k-1} r^{m-1}-{bk\over 2} t^{k-1}r^{m-1} - {am\over 2} t^kr^{m-1} = 0\ ,\nn\\
-k(k-1) t^{k-2}r^m-m(m-1)t^kr^{m-2} + akt^{k-1}r^m + bmt^kr^{m-2} = 0\ , \\
-k(k-1)t^{k-2}r^m + m(m-1)t^kr^{m-2} -2 e^{at}r^b t^kr^m = 0 \ ,\nn\\
{b\over r^2} + 2 e^{at}r^b = 0\ . \nn
\eea
From the last equation, we have $a=0$ for a nontrivial solution
(matching the $t$-exponents): so $b=-2$. From the third
equation, we take $k=1$ for a time-dependent solution. Thus
\be
(m+1)r^{m-1}=0 ,\qquad  (m(m-1)-2)tr^{m-2}=0,\qquad (-m(m-1)+bm) t r^{m-2} = 0\ .
\ee
So finally, we obtain
\be\label{Xf-t-2d}
a=0,\quad\ b=-2,\quad\ k=1,\quad\ m=-1\qquad 
\Rightarrow\qquad X^2={t\over r}\,,\quad\ e^f={1\over r^2}\,.
\ee
The metric is simply $AdS_2$ and the dilaton vanishes at $t=0$,
exhibiting a Big-Crunch \ (while also growing towards the $AdS_2$
boundary at $r\sim 0$).\ In cases where JT arises from some higher
dimensional reduction, the vanishing dilaton at $t=0$ signals a
spacelike singularity where the higher dimensional theory has a Crunch.

It is interesting to compare this with the black hole in JT theory:
in lightcone coordinates $x^\pm=t\pm r$, this is 
\be\label{JT-2dbh}
ds^2 = {-4dx^+dx^-\over (x^+-x^-)^2}\ ,
\qquad\quad  X^2={1-\mu x^+x^-\over x^+-x^-}\ ,
\ee
with $\mu$ the mass. See \eg\ \cite{Almheiri:2014cka,Dhar:2018pii,
  Moitra:2019xoj} for some recent discussions on black hole formation.
We recall that a 4-dim Schwarzschild black hole has a spacelike
Big-Crunch singularity which is Kasner-like (\ref{FlatKas}).
One might then wonder if the Crunch (\ref{Xf-t-2d}) is similar to the
spacelike Crunch singularity in the interior of the black hole above,
where $X^2\ra 0$: this singular locus is
\be
1-\mu x^+x^- = 0\ .
\ee
Here we have $x^-={1\over\mu x^+}$ so the induced metric is\ $ds^2>0$
which is spacelike.

Now looking in the vicinity of the singularity, we parametrize, using
$u,v$ small, as
\be
x^+=x_0^++u ,\ \ \ x^-=x_0^-+v ,\ \ \ 1-\mu x_0^+x_0^-=0\ ,\quad u=t+r,\ \
v=t-r\ .
\ee
Then expanding around $x^+_0=x^-_0={1\over\sqrt{\mu}}$ to linear order
gives
\be
X^2\ \sim\ {1-\mu x_0^+x_0^- - \mu (x_0^+v+x_0^-u)\over x_0^+-x_0^-+u-v}
\ \sim\ {-t\over r}\ .
\ee
This is of the form of the scaling Crunch above, with $t<0$.
Of course instead of looking at the interior of the black hole, we
could take the scaling Crunch to be an independent cosmological
solution in JT gravity, continuing the range of $t, r$: in this case
this violates standard $AdS_2$ boundary conditions where
$X^2\ra {X_0^2\over\epsilon}$ near the boundary $r=\epsilon$.
This in some ways a particular case of the general solution for the
dilaton\ $X^2={1\over r} (a+bt+c(t^2-r^2))$\,.
It is fair to say however that any such solution can be $SL(2,R)$
transformed to the black hole in the form (\ref{JT-2dbh}), under
appropriate conditions. So the behaviour in this case is apparently
less interesting than that in the redux of the higher dimensional
cases as we have seen which have more intricate structure.  If we
relax $AdS_2$ asymptotics by \eg\ allowing departures from the throat
region, possibly driven by extra matter, then we expect more
nontrivial time-dependence to arise. It would be interesting to
explore this further.


\section{Some comments on the dual field theory}

The cosmological backgrounds we have discussed with $AdS_D$
asymptotics ($\theta=0,\ z=1$) are time-dependent deformations of
$AdS/CFT$ \cite{Das:2006dz,Das:2006pw,Awad:2007fj,Awad:2008jf}: these
include $AdS$ Kasner (sec.~2) and other singularities. For $AdS$
Kasner (\ref{AdSDKas}), the dual theory lives on an isotropically
Crunching space $ds^2=-dt^2+t^{2/d_i}dx_i^2$ with a time-dependent
gauge coupling $g_{YM}^2=e^\Phi=t^\al$ where $\al>0$.  Although at
first sight one might imagine the dual to be weakly coupled since the
gauge coupling is vanishingly small at $t\ra 0$ which is the location
of the bulk cosmological singularity, this turns out to not be the
case and interactions are important in general. The coupling is
varying rapidly since $\Phi=\al\log t$ so ${\dot\Phi}^2$ diverges.
Then for instance the gauge kinetic terms ${1\over g_{YM}^2(t)} {\rm
  Tr} F^2$ can be transformed to canonical ones by redefining the
gauge fields (as in standard perturbation theory) by absorbing the
coupling into the definition of the gauge field $A_\mu$. However this
gives rise to new tachyonic, divergent, mass-terms stemming from
time-derivatives of the coupling. These ensure that the field
variables get driven to large values as $t\ra 0$. Retaining the gauge
kinetic terms as above, it turns out that the time-dependent
Schrodinger wavefunctional near $t\ra 0$ has a wildly oscillating
phase, and a divergent amount of energy is pumped in by the external
time dependence \cite{Awad:2008jf}.  This suggests that the gauge
theory response is singular if the coupling strictly vanishes. For
null time-dependence, the redefined gauge field variables (after
absorbing the coupling) in fact lead to the potential mass terms
vanishing due to the lightcone time dependence, thus suggesting weakly
coupled Yang-Mills theory near $x^+\ra 0$\ \cite{Das:2006pw}.  All
these arguments are subject to more detailed analysis of possible
renormalization effects (with a cutoff introduced near the
singularity). For instance it could be that the effective renormalized
coupling does not strictly vanish, so that the dual field theory is
not singular. In this case, the bulk dual might be expected to exhibit
``bounce''-type behaviour. See
\cite{Engelhardt:2015gla,Engelhardt:2016kqb} for related comments on
the dual to a Big-Crunch, based on a no-transmission principle.

We now describe some broader aspects of the time-dependent
deformations of $AdS_5/CFT_4$ in
\cite{Das:2006dz,Das:2006pw,Awad:2007fj,Awad:2008jf}. These
involve the bulk string theory on \eg\ $AdS_5\times S^5$ (in
Poincare slicing) with constant string dilaton $\Phi$, deformed as
\be\label{adsBB}
ds^2 = {R^2\over r^2} ({\tilde g}_{\mu\nu} dx^\mu dx^\nu + dr^2)\ ,
\qquad {\tilde R}_{\mu\nu} = {1\over 2} \del_\mu\Phi\del_\nu\Phi\ ,\qquad
     {1\over\sqrt{-{\tilde g}}}
     \del_\mu\big(\sqrt{-{\tilde g}}\,
         {\tilde g}^{\mu\nu} \del_\nu\big)\Phi=0\ .
\ee
Here ${\tilde g}_{\mu\nu}, \Phi$ are functions of the boundary
coordinates $x^\mu$ alone.  (We are suppressing the 5-sphere part
of the metric $d\Omega_5^2$ as well as the corresponding 5-form RR-flux
supporting these backgrounds.)  In other words, the 4-dim ``boundary''
part of the bulk space is deformed but in a constrained manner,
sourced by a corresponding deformation for the string dilaton $\Phi$.
These constraints arise from requiring that the deformed background
solves the IIB supergravity equations of motion.

With functional dependence restricted to time $t$ or null time $x^+$
alone, we obtain time-dependent deformations of $AdS/CFT$.
If we deform the gauge coupling in the dual SYM to have external
time-dependence as\ $g_{YM}^2=e^\Phi = t^\al$\,, the rapid time variation
as $t\ra 0$ leads to a cosmological singularity in the bulk. Curvature
singularities diverge near $t\ra 0$ as \eg\
\be
R_{tt} \sim {\dot\Phi}^2 \sim {1\over t^2}\ .
\ee
With null time-dependence, curvature invariants are finite but tidal
forces diverge, somewhat similar to \cite{Craps:2005wd}.

The fact that these deformations are constrained has interesting
consequences for the gauge theory. Since we have turned on a
non-normalizable deformation for the metric, we would ordinarily
expect a nonvanishing response, in particular for the stress tensor
which is the dual operator. However incorporating appropriate
counterterms, it turns out that the stress tensor vanishes
\cite{Awad:2007fj}: the holographic stress tensor is
\be\label{Tmunu-AdS}
T_{\mu\nu} = {1\over 8\pi G_5} \Big(K_{\mu\nu} - K h_{\mu\nu}
- 3h_{\mu\nu}  + {1\over 2} G_{\mu\nu} - {1\over 4} \del_\mu\Phi \del_\nu\Phi
+ {1\over 8} h_{\mu\nu} (\del\Phi)^2\Big)\ .
\ee
This is due to cancellations between distinct contributing terms
arising from the metric deformation and the string dilaton scalar
$\Phi$\footnote{For a bulk metric\ $ds^2 = {dr^2\over r^2} +
  h_{\mu\nu} dx^\mu dx^\nu$\ with boundary $r=const$, the outward
  pointing unit normal is\ $n=-{dr\over r}$ and the extrinsic
  curvature is\ $K_{\mu\nu} = -{1\over 2} (\nabla_\mu n_\nu +
  \nabla_\nu n_\mu) = \Gamma_{\mu\nu}^r n_r = {1\over 2} g^{rr}
  (-h_{\mu\nu,r}) n_r = {r\over 2} h_{\mu\nu,r}$. For the backgrounds
  (\ref{adsBB}), the boundary metric is\ $h_{\mu\nu} = {1\over r^2}
        {\tilde g}_{\mu\nu}$, giving\ $K_{\mu\nu} =
        -h_{\mu\nu}$.\ Then the first three terms cancel: the last
        three terms cancel using the constraining relations between
        the deformations.}. This suggests that in some sense, the
deformations are tuned so that the dual response vanishes.
See \cite{Narayan:2012wn} for similar effects, in the context of the
Lifshitz string constructions in \cite{Balasubramanian:2010uk} (which
are related to (\ref{adsBB})).
As discussed there, a smooth metric is in general expected to have
various subleading coefficients nonzero in the Fefferman-Graham
expansion\ of the metric and scalar. Conversely the deformation
$g_{\mu\nu}^{(0)}$ alone will be singular, as is the case above.  In
the present case, it can be seen using the holographic renormalization
prescriptions \cite{de Haro:2000xn,Skenderis:2002wp} that in fact
requiring that the subleading coefficients vanish\footnote{Recall
  the Fefferman-Graham expansion for the metric\
$ds^2 = {dr^2\over r^2} + {1\over r^2} (g^{(0)}_{\mu\nu} + r^2g^{(2)}_{\mu\nu}
+ r^4g^{(4)}_{\mu\nu} + \ldots) dx^\mu dx^\nu$\ and for the scalar\
$\Phi=r^{(d-\Delta)/2} (\Phi^0+r^2\Phi^2+\ldots)$.\
Then solving\ $R_{MN}=-4g_{MN}+{1\over 2}\del_M\phi\del_N\phi$ 
iteratively gives
\be\label{g2constr}
g_{\mu\nu}^2\sim R_{\mu\nu}^0-{1\over 2}\del_\mu\Phi\del_\nu\Phi
-{1\over 2(d-1)}\Big(R-{1\over 2} (\del\Phi)^2\Big) g_{\mu\nu}^0 :\qquad 
g^2_{\mu\nu}=0\ \ \Rightarrow\ \ 
R_{\mu\nu}^0={1\over 2}\del_\mu\Phi\del_\nu\Phi\ ,
\ee
(for a massless scalar $\Delta=d$) also implying the higher order 
coefficients vanish if $g^{(4)}=0$. 
Likewise, with $\Box^0$ being the Laplacian w.r.t. $g^0_{\mu\nu}$, 
we also obtain\ $\Phi^{(2)}\sim \Box^0\Phi^0$: thus $\Phi^{(2)}=0$ 
implies $\Box^0\Phi^0=0$.\ The backgrounds above
thus appear constrained from this point of view, with only the first
coefficient $g^{(0)},\ \Phi^{(0)}$ nonzero for all $r$, the subleading
pieces of the metric and scalar vanishing.}
leads to the constraint conditions (\ref{adsBB}).
It is however worth noting that the above arguments are
screen-dependent: changing screens using Penrose-Brown-Henneaux (PBH)
transformations in fact makes the holographic stress tensor
nonvanishing \cite{Awad:2007fj}.  For instance, for $AdS$ Kasner
(\ref{AdSDKas}), the isotropically Crunching space
$ds^2=-dt^2+t^{2/d_i}dx_i^2$ on which the dual CFT lives is
conformally flat: a PBH transformation transforms this to a new flat
space screen with a nonvanishing holographic stress tensor.

Overall this suggests that the state of the gauge theory dual to the
bulk theory containing cosmological singularities in the screens
(\ref{adsBB}) is some nontrivial, exotic state: for instance severe
time-dependent deformations on the vacuum state would be expected to
thermalize, or equivalently lead to black hole formation in the bulk
\cite{Awad:2007fj}.  Now imagine the situation where RG effects lead
to an effective renormalized coupling that does not strictly vanish:
in this case the boundary theory is likely to be nonsingular, which
suggests the bulk dual undergoes a nonsingular bounce. Related
comments appear in \cite{Engelhardt:2015gla,Engelhardt:2016kqb}, based
on a no-transmission principle.
In this case, the bulk background (\ref{adsBB}) is likely to be
deformed, thus possibly containing subleading terms in the
Fefferman-Graham expansion: then the holographic stress tensor may not
vanish since the cancellations in (\ref{Tmunu-AdS}) may not occur
precisely. It would be interesting to understand this more concretely.

It would appear that several of the above arguments also apply to
deformations of this kind for $AdS_4\times X^7$: in this case the
scalar $\Phi$ must arise from the G-flux in M-theory via
compactification on some appropriate 7-manifold $X^7$. The 11-dim
equations of motion would then lead to analogs of the conditions
(\ref{adsBB}). For ${\tilde g}_{\mu\nu}$ and $\Phi$ being null
deformations, this was discussed in \cite{Balasubramanian:2010uk}. If
this also holds for time-dependent deformations, then presumably
similar comments as above will hold for the dual deformations of the
ABJM theory.


In the JT case (sec.4.3), the Big-Crunch scaling solution
(\ref{Xf-t-2d}) exhibits a vanishing dilaton $X^2\ra 0$ even at the
boundary $r\ra 0$, violating standard $AdS_2$ boundary conditions
with $X^2\sim {X_\epsilon^2\over\epsilon}$ at the boundary
$r=\epsilon$. Turning on time-dependent deformations with standard
$AdS_2$ boundary conditions is expected to lead to black hole
formation\ (see \eg\ \cite{Almheiri:2014cka,Dhar:2018pii,Moitra:2019xoj}
which studies black hole formation in 2-dim dilaton gravity). Thus
the Big-Crunch case we have discussed in JT gravity also appears to be
non-generic. Perhaps correlation functions, OTOCs, string probes (see
\eg\ \cite{Banerjee:2018kwy}) and so on will help in making more
precise the signatures of the Big-Crunch here, in comparison with
black holes.

The 2-dimensional backgrounds from compactifications that we have
discussed are a subclass of (\ref{adsBB}) compatible with the
KK-reduction ansatz (\ref{ads2Redux1}). The analogs of the constraint
conditions from the intrinsically 2-dim point of view would appear to
translate on constraints on the exponents, through the equations of
motion (which are in fact what lead to the constraints
(\ref{adsBB})). Perhaps this is related to the universal near singular
behaviour and the space of exponents being so constrained.

\section{Discussion}

We have argued for universal behaviour near Big-Bang or -Crunch
singularities in various classes of theories, including flat space
(\ref{FlatKas}), $AdS$ (\ref{AdSD-exp2}), (\ref{AdSKas-XfPhi}),
conformally $AdS$ (\ref{Nconf-t}) and more general nonrelativistic
theories with nontrivial Lifshitz scaling and hyperscaling violation
(\ref{hvLif-t}). The near singularity region exhibits universal
interrelations in the behaviour in time.  These are essentially
analogs of the familiar Kasner singularity with minimal anisotropy:
embedding into various asymptotic regions gives new features
however. As we have seen, it is convenient to analyse this in terms of
an effective 2-dimensional dilaton-gravity-scalar theory
(\ref{actionXPhiU}), performing dimensional reduction as
(\ref{ads2Redux1}), (\ref{weyl}): this then reveals the near
singularity behaviour in the form (\ref{nearSing-EMD-1}),
(\ref{univSing}), where the dilaton potential encoding information
about the asymptotic data defined by the theory has disappeared.  (It
is worth mentioning that the effective 2-dim analysis also helps
recover various time-independent backgrounds.) This is by no means an
exhaustive classification: for instance our detailed analysis based on
the scaling ansatz (\ref{XfPhi-ansatz}) relied on the apparent
separability of our equations of motion in time and space, leaving
open the question of backgrounds that do not admit such factorization.

From the 2-dimensional point of view, the theories here are more
complicated than Jackiw-Teitelboim, in particular with an extra scalar
and a nontrivial dilaton potential playing essential roles in driving
near singularity dynamics. We have mostly used the 2-dim theory as a
crutch that faithfully captures the higher dimensional
theories. Looking at the spatial profile of the dilaton $X$, we see
that $X^2\sim r^m$ with $m$ reflecting a non-normalizable mode in the
higher dimensional theory: for instance for the $AdS_D$ deformation
(\ref{AdSD-exp1}) comprises a non-normalizable mode in $D=d_i+2$
dimensions, and so is distinct from one in 2-dimensions
intrinsically. Likewise the spacetime metric singularities are
reflected in higher dimensional curvature invariants
(\eg\ $R_{ABCD}R^{ABCD}$) diverging.  However the 2-dim cosmological
singularities are interesting in their own right: it would be
interesting to understand them better, perhaps as (exotic)
deformations of $AdS_2$, possibly using various insights from
investigations of the SYK and related models and nearly $AdS_2$
holography
\eg\ \cite{Almheiri:2014cka,Maldacena:2016upp,Jensen:2016pah,
  Engelsoy:2016xyb,Sarosi:2017ykf,Rosenhaus:2018dtp,Trunin:2020vwy}. See \eg\
\cite{Witten:2020ert} for certain classes of deformations of JT gravity.

The scaling ansatze (\ref{XfPhi-ansatz}) we have considered 
\be\label{XfPhi-ansatz-Dis}
X^2=t^kr^m ,\ \ \ e^f=t^ar^b ,\ \ \ e^\Phi=t^\al r^\beta\quad\ra\quad
ds^2 = {e^f\over X^{2(d_i-1)/d_i}}\big(-dt^2+dr^2\big)+X^{4/d_i}dx_i^2\ ,\ \ 
\ee
are designed to simulate a Big-Crunch (or -Bang), with the various fields
becoming vanishingly small at some instant of time ($t=0$): so they
are best thought of as near singularity ansatze.
Through our analysis these lead to the universal interrelations
$k=1,\ a={\al^2\over 2}$\,, in the time-dependent behaviour, with the
precise values of $\al$ determined by the asymptotic data. The near
singular bulk region in the higher dimensional theory takes the form
\be
ds^2 = {r^{2\theta/d_i}\over r^2} \left( {1\over t^{\gamma\al}}
\Big(-{dt^2\over r^{2z}}+dr^2\Big) + t^{2/d_i} dx_i^2\right) ,\qquad
e^\Phi=t^\al r^\beta\ .
\ee
For $\theta=0, z=1$, we have $\gamma=0, \beta=0$: these are $AdS$
Kasner singularities sourced by $\Phi$.
Note that in most of the cases we have studied, the exponent $\al$
governing the time-dependence of $e^\Phi$ is irrational: this stems
from the fact that $\al$ is determined by a quadratic
(\ref{bbetamaal-2}). This renders tricky the possibility of
analytically continuing $e^\Phi$ from early times $t<0$ across the
singularity at $t=0$ to late times $t>0$. However an interesting
exception is the $AdS_4$ Big-Crunch (\ref{ads4bb}): this has
$d_i=2,\ \theta=0$ so (\ref{bbetamaal-2}) does not give irrational
$\al$. These are deformations of $AdS_4\times X^7$, the scalar $\Phi$
in this case arising from the 4-form flux in M-theory after
compactification on some 7-manifold $X^7$: the duals are expected to
be deformations of the ABJM theory.

Although the near singularity behaviour exhibits universality in terms
of universal interrelations \emph{between} the exponents, the precise
values of the exponents are determined by the full theory. In
particular this depends on the theory-dependent dilaton potential into
which the singular region is embedded (and as we have seen, the
space of exponents is quite constrained). The detailed form of the
cosmological solution containing the Big-Crunch singularity, as well
as its existence, appear to depend intricately on the asymptotic
region far from the singularity. This is a bit reminiscent of the
UV-IR mixing discussed in \cite{Minwalla:1999px}, here in the sense
that near-singularity or short-time physics intertwines with long-time
physics. Note however that the analysis here is entirely classical
gravity: this suggests that any such UV-IR mixing at play here is
encoded within the gravity approximation. In some essential sense, the
deformations are ``constrained'' in the sense described in sec.5,
suggesting that the CFT state dual to such a cosmological singularity
is some nontrivial, exotic, state, in some ways encoding information
about the existence of the singularity, which perhaps reflects the
UV-IR mixing above. By comparison, generic time-dependent deformations
on generic CFT states would be expected to thermalize (equivalently,
lead to black hole formation). It would be interesting to understand
this better.  In this regard, it is also worth noting that some of
these geometries encode holographic flow in both space and time:
\eg\ the cosmological nonconformal brane backgrounds (\ref{Nconf-t})
for the D2-brane case ($d_i=2$) might be regarded as encoding a
renormalization group flow towards the M2-brane $AdS_4$ Kasner
singularity (\ref{AdSKas-XfPhi}), noting that the time-independent
versions indeed encode such an RG flow encapsulated by the D2-M2 brane
phase diagram \cite{Itzhaki:1998dd}. In this context, see the recent
paper \cite{Frenkel:2020ysx}: the backgrounds there are consistent
with the dimensional reduction ansatze we have discussed, and may be
instructive in analysing more general flows.

It is worth noting that the scaling ansatz (\ref{XfPhi-ansatz-Dis})
above reflects a singular Big-Crunch (or -Bang) at say $t=0$: as we
have seen, this is quite constraining and leads via the detailed
analysis to the precise values we have discussed. However we could
relax the requirement of the fields actually Crunching fully: if
instead we allow for bounce-type behaviour (with
\eg\ $X^2\xrightarrow{t\ra 0} X_0^2\neq 0$ and so on), then one might
imagine more general possibilities. It would be interesting to explore
this further.

We have retained isotropy in the spatial $x_i$ directions: this is
consistent with the reduction to 2-dimensions. Introducing anisotropy
here will of course enlarge the space of such cosmologies.  It is
likely that performing the kind of analysis we have done for fully
anisotropic ansatze will reveal analogs of the more general
anisotropic Kasner singularities embedded in various asymptotic
structures: it would seem that these must exist, but it would be
interesting to understand this existence question conclusively. More
ambitiously it would be interesting to understand more general
homogenous spaces as in the Bianchi classification to understand
analogs of BKL-type singularities  \cite{Landau}-\cite{Damour:2002et}
for more general asymptotic structures (see \cite{Awad:2008jf}
for the $AdS$ case).

From the dual point of view, the universal near singularity bulk
behaviour suggests a universal dual time-dependent strongly coupled
large $N$ matrix quantum mechanics, comprising the spatial reduction
of a CFT on a time-dependent base space subjected to a severe
time-dependent gauge coupling. For instance the duals to the
reductions to 2-dimensions of the $AdS_5\times S^5$ deformations are
expected to arise from the spatial $T^3$ reduction of the
time-dependent deformed dual $CFT_4$. One might imagine that this dual
is some core subsector of the dual CFT possibly encoding some
essential features of the bulk singularity.  It appears unlikely
however that this decouples from the rest of the theory: the dual is
instead perhaps better thought of as the full dual theory
dimensionally reduced on the compact space. We hope this point of view
is of interest however in understanding the duals to cosmological
singularities. In particular it would be of interest to understand
possible signatures of the universal behaviour on correlation
functions, OTOCs and other observables. We hope to explore this in
future work.

Finally, it is interesting to recall that JT gravity was argued to be
nonperturbatively dual to a random matrix model \cite{Saad:2019lba}
(see also \cite{Stanford:2019vob}): this gravity/ensemble duality
dovetails with the existence of replica wormholes
\cite{Penington:2019kki,Almheiri:2019qdq} (see \cite{Almheiri:2020cfm}
for a review of some aspects of these issues).  More generally this
raises the question of the duals to 2-dim dilaton gravity theories
being ensembles of various theories, rather than specific unitary
theories (see \eg\ \cite{Marolf:2020xie,Witten:2020ert} for related
discussions in more general theories). In all our discussions, we have
been mostly regarding the 2-dim dilaton gravity theories as crutches
for the higher dimensional theories, which we expect are dual to
specific theories (see \cite{McNamara:2020uza} for some discussions in
this regard). Also, the 2-dim theories here have substantially more
complicated dynamics than JT gravity, faithfully encoding properties
of the higher dimensional theories that they are compactifications of:
this suggests their behaviour is likely to be fundamentally different
from that of JT gravity. It would be fascinating to understand with
greater clarity if the duals to the 2-dim dilaton gravity theories
here are specific theories or ensembles.


\vspace{10mm}

{\footnotesize \noindent {\bf Acknowledgements:}\ \ It is a pleasure
  to thank Daniel Grumiller for an interesting early discussion on
  aspects of 2-dimensional dilaton gravity from reduction, and Sumit
  Das, Dileep Jatkar, Arnab Kundu, Shiraz Minwalla, Sandip Trivedi and
  Amitabh Virmani for discussions and comments on a draft. KN also
  thanks Sumit Das, Shiraz Minwalla and Sandip Trivedi for many
  discussions over the years on $AdS$ cosmologies and their
  duals. This work is partially supported by a grant to CMI from the
  Infosys Foundation.  }


\end{document}